\documentclass{ws-mpla}
\usepackage[super]{cite}
\usepackage{graphicx}
\usepackage[utf8x]{inputenc}
\usepackage[english]{babel}
\usepackage{graphicx}
\usepackage{float}
\usepackage{multirow}
\usepackage{xcolor}
\usepackage{cite}
\usepackage{setspace, longtable, amstext}
\usepackage{selinput}
\usepackage{hyperref}

\usepackage{amssymb}
\usepackage{amsmath}

\usepackage{amsopn}
\def\beq{\begin{equation}}
\def\eeq{\end{equation}}
\def\bea{\begin{eqnarray}}
\def\eea{\end{eqnarray}}

\begin{document}

\markboth{Authors' Names}
{Instructions for Typing Manuscripts (Paper's Title)}

\catchline{}{}{}{}{}

\title{On the role of dust in the microwave emission of galactic halos}

\author{A. Amekhyan} 
\author{S. Sargsyan} 
\author{A. Stepanian}
\address{Center for Cosmology and Astrophysics, Alikhanian National Laboratory, Yerevan, Armenia
}

\maketitle

\begin{history}
\received{... 2019}
\accepted{...}
\end{history}

\begin{abstract}
The contribution of the thermal dust component in galactic halo rotation is explored based on the microwave data of Planck satellite. The temperature asymmetry of Doppler nature revealed for several edge-on galaxies at several microwave frequencies is analyzed regarding the contribution of the thermal dust emission. We derive the dust contribution to the galactic halo rotation using the data in three bands, $353 \mathrm{GHz}$, $545 \mathrm{GHz}$ and $857 \mathrm{GHz}$ for two nearby galaxies M81 and M82. The relevance of the revealed properties on the halo rotation is then discussed in the context of the modified gravity theories proposed to describe the dark matter configurations. 
\end{abstract}


\ccode{PACS numbers: 98.62.Gq}

\section{Introduction}	

The study of the Planck data \cite{Pl} in several microwave bands had revealed temperature asymmetry for several nearby edge-on spiral galaxies extending to their halos \cite{DP14,DP15,DP16,G15,G18}. That asymmetry is frequency independent, thus indicating its Doppler nature.
Previously we have studied \cite{Amekhyan} that effect based on the microwave data for the galaxy M31, namely, regarding the possible contribution of the thermal dust in the rotation of the halo. Here we continue the analysis, now using the data on galaxies M81 and M82. Since the galactic halos act as probes for testing of modified gravity models, we therefore discuss also that aspect with certain models  based on revealed role of the thermal dust in the rotation of the halos. 

Among various models the baryonic matter in the form of molecular clouds was suggested as the possible content of the dark matter in galaxies including their halos \cite{DP95}. Then, the mentioned microwave asymmetry of Doppler nature has to reflect that. In our previous study of M31 galaxy \cite{Amekhyan} using the microwave temperature asymmetry data we have concluded that the thermal dust cannot be the main contributor to the dark matter, since the rotation velocity provided by dust component is quite low. So, here we will perform the analysis for the M82 and M81 galaxies.

Crucial parameter of the interstellar dust is the optical depth, which indicates the fraction of the intensity decreased by extinction of the light emitted by dust particles grains at line-to-sight propagation. In general, when the light ray propagates through a dusty medium it weakens by extinction and is reinforced by thermal emission or scattering. It is clear, that galactic halos are optically thin ($\tau\ll1$) and the optical depth decreases at large galactocentric distances. We obtain the optical depth as a function of the galactocentric radii, for which we use several Planck maps at different frequency bands. As in \cite{Amekhyan}, also here we adopt a Modified Blackbody (MBB) spectrum for deriving optical depth at large distances, as well as two dust models, i.e. DL07 \cite{DL07} and GNILC, see \cite{Aghanim}. The temperature is a key parameter of the dust through which the dust is classified to warm and cold components. The warm dust can absorb energetic photons and heat up to $40\mathrm{K}$. For the cold dust illuminated by the background interstellar radiation field, the temperature is $10-25\mathrm{K}$. In fact, the dust temperature strongly depends on the grain size, chemical composition, distance from the radiation source. So for cold grains, which are in local thermodynamic equilibrium (LTE) with interstellar radiation field (ISRF), the Kirchhoff law is applicable, according to which the ratio of emission and absorption coefficients are given via Planck function. The latter depends only on the temperature for given frequency band. In our work the calculations have been performed for large galactocentric distances corresponding to the presence of ISRF, see \cite{DL07}. The dust temperature we fix from the Planck maps via MBB fitting procedure. Since the major fraction of the interstellar dust consists of cold dust component (over $90\%$ \cite{Suzuki}), we consider cold dust models ($T<22\mathrm{K}$) with typical values of the spectral index $\beta=(1.5-2)$ taken from Planck maps, using the same frequency bands both for cold dust emission and temperature asymmetry maps.

We use Planck High Frequency Instrument (HFI) data \cite{Adam} at three bands, $857 \mathrm{GHz}$, $545 \mathrm{GHz}$, $353 \mathrm{GHz}$ and determine the relevant microwave temperature asymmetry. We have calculated the temperature asymmetry values using the methods of \cite{DP14,DP15,DP16,G15,G18}.

The paper is organized as follows: Sections 2 and 3 describe dust emission properties and velocities for M82, M81 galaxies, respectively. In Section 4 we briefly discuss certain modified gravity models proposed to explain the dark matter paradigm including the  model in \cite{VG, GS1}. In Section 5 we draw our main conclusions.

\section{Rotation velocities of M82 galaxy halo}

M82 is one of the brightest infrared objects located at distance of $3.63\mathrm{Mpc}$ from us (in M81 group)  \cite{Freedman}. It is a starburst  late type galaxy, with  $80^{0}$ inclination angle \cite{Blackman}. The origin of the  nuclear starburst region (with $500\mathrm{pc}$ size\cite{Forster}), is assumed to be due to a close encounter between M82 and NGC3077\cite{Yun}. The near infrared data indicate the presence of bar with $1\mathrm{kpc}$ size \cite{Telesco}. As reported by \cite{Sofue}, the major part of galaxy's mass ($\approx10^{10}\mathrm{M_{\odot}}$) is located in the inner region (within $2\mathrm{kpc}$). Besides this, there is a bulge component with $10^{7}\mathrm{M_{\odot}}$ mass and $7.5\mathrm{pc}$ size \cite{Gaffney}. It also has spiral arms which are not visible on the optical images (due to presence of dust particles). However, they can be seen on the near infrared wavelengths \cite{Mayya}. Also, M82 is a gas rich galaxy with $30-40\%$ gas fraction\cite{Young}. Furthermore, there are some evidences for the presence of dust in the inner disk \cite{YDMayya}. The rotation curve of the M82 has been studied by many authors. According to \cite{Greco}  it is flat within $1-4\mathrm{kpc}$, with $\approx10^{10}\mathrm{M_{\odot}}$ dynamical mass. However, starting from $1.5\mathrm{kpc}$ to $10\mathrm{kpc}$ radius, the rotation velocity, derived by HI kinematics, sharply decreases from $120\mathrm{km/s}$ to $50\mathrm{km/s}$ \cite{Martini}.
It should be mentioned that although different components of M82, such as bar, bulge, disk or spiral arms are well studied, less information is available about its halo. 

\textit{2.1 Dust emission parameters for M82 halo}:
Here we attempt to obtain radial behavior of dust emission optical depth using several Planck maps \cite{Abergel, Ade}. 
The description of used Planck maps, as well as calculation technique is described in \cite{Amekhyan}. Similarly to M31 \cite{Amekhyan}, we calculate optical depth for two dust models: DL07\cite{DL07} and GNILC\cite{Aghanim}. The necessary parameters of the MBB spectrum, dust spectral index $\beta$ and temperature $T_{d}$, are taken from Planck GNILC maps. The parameters used in this paper are calculated as follows: from Planck map we take circles of $15\mathrm{kpc}$, $20\mathrm{kpc}$ and $25\mathrm{kpc}$ radii, their centers coincide with the center of the galaxy. Then we calculate the average value of each parameter for given radii.  For DL07 model spectral index is equal to 2 for all radii, while in the case of GNILC model  $\beta$ varies within large distances. For M82 and M81 galaxies there is also an anti correlation between $\beta$ and $T_{d}$ (see Table 1), which is appeared during MBB spectrum fitting (caused by noise) \cite{Shetty}.
\begin{table}[h!]   
\tbl{Dust temperature, spectral index and microwave asymmetry data.}
{
\centering
\begin{tabular}{ |p{1.2cm}||p{1.5cm}| |p{1.5cm}||p{1.5cm}| |p{2.3cm}||p{2.5cm}| | }
\hline
 Distance&\multicolumn{3}{|c|}{Microwave temperature asymmetry $\frac{|\Delta T|}{T_{d}}$($\mu K$)}&\multirow{2}{*}{Spectral index ($\beta$)}&\multirow{2}{*}{Temperature $T_{d} (K)$}\\
 \cline{2-4}
  $r (kpc)$ & 857 GHz &545 GHz &353 GHz& & \\
\hline
   15&0.0493&0.305&4.81 &1.577&20.381\\
   20&0.0472&0.305&3.24&1.639&19.382\\
   25&0.0396&0.254&1.68&1.687&18.695\\
  \hline
  \end{tabular}
}
\end{table}

As already mentioned, we derive  optical depth $\tau$ from MBB spectrum. It should be noticed that $\tau$  remains almost unchanged for $T_{d}\neq const$, however, it strongly depends on the spectral index. For example, in the case of DL07 ($\beta=2$) the order of $\tau$ is 10$^{-3}$, while for GNILC ($\beta\neq const$) it is 10$^{-2}$ . In fact, for all models and all cases $\tau$ decreases for large radii (see Tables 2-4).
 \begin{table}[h!]   
\tbl{The optical depth for DL07 model.} 
{
\centering
\begin{tabular}{ |p{2cm}||p{2cm}| |p{2cm}||p{2cm}| }
 \hline
  \multicolumn{4}{|c|}{DL07, $\beta=2,$ $T=19.48K$} \\
 \hline
 Distance\newline r (kpc)
 
 &Optical depth\newline$\tau_{857}$ $(10^{-3})$ 
 
 &Optical depth\newline$\tau_{545}$ $(10^{-3})$
 
 &Optical depth\newline$\tau_{353}$ $(10^{-3})$ \\
 
  \hline
  15&$1.730$&$2.959$&$4.602$\\
 
  20&$1.231$&$2.306$&$3.431$\\
 
  25&$1.038$&$1.982$&$3.210$\\
  \hline
  \hline
\multicolumn{4}{|c|}{DL07,  $\beta=2,$ $T\neq const$}\\ 
   \hline
Distance\newline r (kpc)

&Optical depth\newline $\tau_{857}$ $(10^{-3})$     
 
&Optical depth\newline $\tau_{545}$ $(10^{-3})$

&Optical depth\newline $\tau_{353}$ $(10^{-3})$\\

  \hline

   15&$1.650$&$2.818$&$4.387$\\
 
   20&$1.231$&$2.313$&$3.515$\\
  
	 25&$1.079$&$2.064$&$3.344$\\
 
	\hline
  \end{tabular} \label{tab:2}}
	
\end{table}
\begin{table}[h!]   
\tbl{ The optical depth for GNILC model with constant $\beta$.}
{
\centering
\begin{tabular}{ |p{2cm}||p{2cm}| |p{2cm}||p{2cm}| }
 \hline
  \multicolumn{4}{|c|}{GNILC,   $\beta=2$, $T\neq const$} \\
  \hline
 Distance\newline r (kpc)
 &Optical depth\newline$\tau_{857}$ $(10^{-3})$
 &Optical depth\newline$\tau_{545}$ $(10^{-3})$
 &Optical depth\newline$\tau_{353}$ $(10^{-3})$\\
  \hline
    
   15&1.065&2.492&4.341\\

   20&0.973&2.295&4.058\\
 
   25&0.941&2.160&3.721\\
  \hline
  \end{tabular} \label{table:3}}
 \end{table}

\begin{table}[h!]    
\tbl{The optical depth for GNILC model with varying $\beta$.}
{
\centering 
\begin{tabular}{ |p{2cm}||p{2cm}||p{2cm}||p{2cm}|  }
\hline
\multicolumn{4}{|c|}{GNILC, $\beta\neq const$, $T=const$} \\
\hline
Distance\newline r (kpc)
&Optical depth\newline $\tau_{857}$ $(10^{-2})$
&Optical depth\newline $\tau_{545}$ $(10^{-2})$ 
&Optical depth\newline $\tau_{353}$ $(10^{-2})$\\
\hline
  15&1.858&3.584&4.932\\
  20&1.116&2.223&3.366\\
  25&0.778&1.551&2.422\\
\hline
\end{tabular} \label{table:4}
}
\end{table}

\textit{2.2 M82 dust rotation velocities}:  The rotation velocities of M82 are obtained via the formula \cite{DP95} 
\begin{equation}\label{V}
\frac{|\Delta T|}{T_{d}}=\frac{2v\, sin i}{c}\tau,
\end{equation} 
where the inclination angle $i=80^{^{\circ}}$, the temperature asymmetry $\frac{|\Delta T|}{T_{d}}$ and optical depth $\tau$ depend not only on the radius, but also on the given frequency band. In Eq.(1) $v$ is the dust rotational velocities at $15\mathrm{kpc}$, $20\mathrm{kpc}$, $25\mathrm{kpc}$ distances. So, based on Eq.(1) we aim to find out whether interstellar dust component can give a significant contribution in the halo rotation. Since we use two different dust models at three frequency bands, so the obtained velocities are different.  The temperature asymmetry has relatively high values especially at $353 \mathrm{GHz}$ band (see Table 1). As we can see from Table 5 and 6, the rotation velocities vary significantly  according to frequency. On the other hand velocities derived by DL07 dust model, have their lowest values at $857 \mathrm{GHz}$ band. In this case, there is no significant change of velocity within the specified radius. Namely, at $545 \mathrm{GHz}$ band it rises up to $20\mathrm{kpc}$, then decreases only by $0.61\mathrm{km/s}$. Finally, at $353 \mathrm{GHz}$ it decreases from $159\mathrm{km/s}$ to $80\mathrm{km/s}$. As in the DL07 case, for GNILC the velocities have their lowest values at $857 \mathrm{GHz}$ and relatively higher ones at $545 \mathrm{GHz}$ and $353 \mathrm{GHz}$ bands.
\begin{table}[h!]   
\tbl{M82 dust rotational velocities by DL07 model.}
{
\centering
\begin{tabular}{ |p{2cm}||p{2cm}||p{2cm}||p{2cm}|}
\hline
\multicolumn{4}{|c|}{DL07, $\beta=const$,   $T=const$} \\
\hline
Distance\newline r (kpc)
&$\nu=857$ (GHz)\newline  $V_{rot}$ (km/s)
&$\nu=545$ (GHz)\newline  $V_{rot}$ (km/s)
&$\nu=353$ (GHz)\newline  $V_{rot}$ (km/s)\\
\hline
  15& 4.35&15.76&159.69 \\

  20&5.86& 20.24&163.74 \\
 
  25& 5.84 & 19.63&80.12 \\

\hline
\hline

  \end{tabular} \label{table:7}}

\end{table}

\begin{table}[h!]   
\tbl{M82 dust rotational velocities by GNILC model.}
{
\centering
\begin{tabular}{ |p{2cm}||p{2cm}||p{2cm}||p{2cm}|}
\hline
\multicolumn{4}{|c|}{GNILC $\beta\neq const$,   $T=const$} \\
\hline
 
Distance\newline r (kpc)
&$\nu=857$ (GHz) \newline  $V_{rot}$ (km/s)
&$\nu=545$ (GHz)\newline  $V_{rot}$ (km/s)
&$\nu=353$ (GHz)\newline  $V_{rot}$ (km/s)\\ 
  \hline
   15& 0.405& 1.302 & 14.91 \\

   20&0.647 & 2.10 &14.71 \\
 
   25& 0.779 & 2.507  &10.61 \\
 
\hline
\hline
\end{tabular} \label{table:6}}
\end{table}
The optical depth obtained according to MBB formula strongly depends on the flux density in given frequency band and the radial distance. Flux density decreases at large galactocentric distances. It also decreases from $857\mathrm{GHz}$ to $353\mathrm{GHz}$ band. In general, the dust radiates on broad range of frequencies, e.g. the warm grains, which have small size, emit mainly at NIR/MIR range, while cold grains of larger size emit at submillimeter range.  The cold dust within the ISRF emits at high frequencies ($\nu>353\mathrm{GHz}$), however for our work we consider the three bands in view of the temperature asymmetry data. Namely, for both - the microwave data and the dust - we consider the same bands. Consequently, at $857\mathrm{GHz}$, the dust rotational velocity has the lowest value. We will discuss these results in more details in Section 2.3 and check that the dust mass has higher values at $353\mathrm{GHz}$, $545\mathrm{GHz}$ and therefore the velocities at these bands are higher than those at $857\mathrm{GHz}$. Actually, the frequency-dependent velocity appears during the calculation of optical depth, since we use the MBB spectrum. Namely, we calculate the mean flux density ($S(\nu)$) for given galactocentric radii and given frequency, then obtain the optical depth $\tau (\nu)$ via MBB formula and finally, according to Eq.(1) we obtain the dust rotational velocity, which depends not only on the dust model, but also on given frequency. 

\textit{2.3 Dust masses}: Dust mass depends on the frequency at which it is emitted. For example, in \cite{Nikola} the dust masses have been estimated at $37\mathrm{\mu m}$ ($\approx8102 \mathrm{GHz}$) and $31\mathrm{\mu m}$ ($\approx9670 \mathrm{GHz}$) within $1\mathrm{kpc}$. According to emission peaks at these wavelengths the estimated masses are $3.6\times 10^{4}\mathrm{M_{\odot}}$ and $1.8\times 10^{4}\mathrm{M_{\odot}}$, respectively. At wavelength ($\lambda =1.2\mathrm{mm}$) \cite{Thuma} the total dust mass in the inner region ($3\mathrm{kpc}$) is $7.5\times 10^{6}\mathrm{M_{\odot}}$. Furthermore, the dust mass has been estimated to be roughly $(1-3)\times 10^{5}\mathrm{M_{\odot}}$ based on the giant star population and starburst model \cite{Hughes} (within $500\mathrm{pc}$). At $450\mathrm{\mu m}$ range ($666 \mathrm{GHz}$), $M_{d}=3.66\times 10^{6}\mathrm{M_{\odot}}$ has been reported by \cite{Hughes94}.  

For our bandwidths we estimate dust mass from \cite{Hil} 
\begin{equation}\label{H}M_{dust}=\frac{\tau D^{2}\Omega}{k}, 
\end{equation} 
where $\tau$, $k=\frac{3Q_{\nu}}{4a \rho}$, $a$ and $\rho$ stand for the optical depth, the mass absorption coefficient, dust grain radius and density, respectively, $\Omega$ is the solid angle, $D$ is the distance between the source and the observer, $Q_{\nu}$ is the grain emission efficiency. Since these quantities are not well defined for our frequencies, we adopt a power law shape of $k$ and parametrized form $k=0.1\mathrm{{cm}^{2}g^{-1}}( \frac{\nu}{1000 \mathrm{(GHz)}})^\beta$ according to \cite{Heithausen}. In this way we estimate the dust mass for $857 \mathrm{GHz}$, $545 \mathrm{GHz}$ and $353 \mathrm{GHz}$ frequencies and two models: DL07 and GNILC (see Tables 7 and 8).

\begin{table}[h!]   
\tbl{M82 dust masses for DL07 model.}
{
\centering
\begin{tabular}{ |p{2cm}||p{2cm}||p{2cm}||p{2cm}|}
 \hline
  \multicolumn{4}{|c|}{DL07, $\beta=const$,   $T=const$} \\
 \hline
Distance\newline r (kpc)
&$\nu=857$ GHz\newline $M_{\odot}$ $(10^{7})$
&$\nu=545$ GHz\newline $(M_{\odot}$ $(10^{7})$
&$\nu=353$ GHz\newline $M_{\odot}$ $(10^{7})$\\ 
\hline
   15& 0.313& 1.265 & 3.542 \\

   20&0.188 & 0.785 &2.417 \\
 
   25& 0.131 & 0.547 &1.739 \\
 \hline
\hline
\end{tabular} \label{table:8}}
\end{table}

\begin{table}[h!]   
\tbl{M82 dust masses for GNILC model.}
{
\centering
\begin{tabular}{ |p{2cm}||p{2cm}||p{2cm}||p{2cm}|}
 \hline
  \multicolumn{4}{|c|}{GNILC, $\beta\neq const$,   $T=const$} \\
 \hline
Distance\newline r (kpc)
&$\nu=857$ GHz\newline  $M_{\odot}$ $(10^{6})$
&$\nu=545$ GHz\newline  $M_{\odot}$ $(10^{6})$
&$\nu=353$ GHz\newline  $M_{\odot}$ $(10^{6})$\\ 
  \hline
   15& 0.291& 1.045 & 3.305 \\

   20&0.207 & 0.814 &2.607 \\
 
   25& 0.174 & 0.700  &2.305 \\

  \hline
\hline
 
  \end{tabular} \label{table:X}}
\end{table}

In fact, even in the same frequencies different dust masses (up to an order of magnitude) are derived from DL07 and GNILC models \footnote{The reason of such differences for the dust masses (in the same frequency) arises from the given dust model i.e. DL07 and GNILC models have different parameters (e.g. spectral index $\beta$, optical depth $\tau$) describing the dust emission properties and composition.}. From these masses one can obtain the dust orbital velocities
\begin{equation}
V^{2}(r)=\frac{GM_{d}}{r},
\end{equation}
where $M_{d}$ is the dust mass. Note that, the dust mass can be minor fraction of the entire galactic dynamical mass and therefore the rotation velocity obtained by the dust mass can define not the global galactic rotational velocity. Regarding the Doppler induced microwave temperature asymmetry, we assume that it is determined by presence of cold dust and hence the obtained velocities are attributed only to the dust component. If in Eq.(3) we also add the stellar mass or other baryonic  components, we will obtain higher velocities, than what we have according to Eq.(1). This is due to the fact that the ISM fraction with respect to the total baryonic mass is about 10$\%$ and the dust mass contribution in ISM is small. On the other hand, by including other baryonic components in the Eq.(3) one has to take them into account also in Eq.(1), since the rotation determined by the microwave temperature asymmetry with respect to dust component is also present there (expressed via dust emission optical depth ($\tau$)).  Since we are interested in the rotation at large galactocentric distances, we do not take into account the effects important at small distances from the galactic center.  The results are presented in Tables 9 and 10.
\begin{table}[h!]   
\tbl{M82 dust rotational velocities by DL07 model.}
{
\centering
\begin{tabular}{ |p{2cm}||p{2cm}||p{2cm}||p{2cm}|}
 \hline
  \multicolumn{4}{|c|}{DL07, $\beta=const$,   $T=const$} \\
 \hline
Distance\newline r (kpc)
&$\nu=857$ GHz\newline  $V_{rot}$ (km/s)
&$\nu=545$ GHz\newline  $V_{rot}$ (km/s)
&$\nu=353$ GHz \newline  $V_{rot}$ (km/s)\\ 
  \hline
   15& 0.950& 1.911& 3.198 \\

   20&0.736 & 1.505 &2.641\\
 
   25& 0.615 & 1.256  &2.240 \\

  \hline
\hline
 
  \end{tabular} \label{table:8}}
\end{table}

\begin{table}[h!]   
\tbl{M82 dust rotational velocities by GNILC model.}
{
\centering
\begin{tabular}{ |p{2cm}||p{2cm}||p{2cm}||p{2cm}|}
 \hline
  \multicolumn{4}{|c|}{GNILC, $\beta\neq const$,   $T=const$} \\
 \hline
Distance\newline r (kpc)
&$\nu=857$ GHz\newline  $V_{rot}$ (km/s)
&$\nu=545$ GHz\newline  $V_{rot}$ (km/s)
&$\nu=353$ GHz\newline  $V_{rot}$ (km/s)\\ 
  \hline
   15& 0.289& 0.549 & 0.976 \\

   20&0.244 & 0.484 &0.867 \\
 
   25& 0.224 & 0.449  &0.815 \\

  \hline
\hline
 
  \end{tabular} \label{table:8}}
\end{table}
Comparing the above results with the velocities obtained with Eq.(\ref{V}), we see that the former have significantly lower values. 

\section{Rotation velocities of M81 galaxy}

M81 (NGC3031) is a SA(s)ab galaxy with $26.9 \mathrm{arcmin}$ angular diameter and about $3.6\pm0.2\mathrm{Mpc}$ distance from us\cite{Gerke}. It is the largest member of the M81 Group  and its apparent magnitude in the B band is 7.69. It has $-35\mathrm{km/s}$ heliocentric radial velocity and its extinction in the B band ($A_{B}$) is 0.36 \cite{Karachentsev}. Also there is an evidence for the presence of HI bridge and interaction  between M81 and M82\cite{Yun}. 

\textit{3.1 M81 dust emission parameters}: We calculate dust emission optical depth using the value of spectral index and dust temperature. According to Table 11, the cold dust temperature is lower compared to M82. Actually, for M82 mean temperature and spectral index \footnote{The values of spectral index for both M81 and M82 galaxies and also the temperature and flux density for each frequency have been calculated from Planck GNILC maps\cite {Aghanim}. Description of the Planck maps, as well as the calculation technique have been presented in \cite{Amekhyan}.} are $T_{d}=19.48\mathrm{K}$, $\beta=1.634$, respectively, whereas for M81 $T_{d}=16.92\mathrm{K}$, $\beta=1.840$. 
\begin{table}[h!]   
\tbl{M81 dust temperature, spectral index and microwave temperature asymmetry data.} 
{
\centering
\begin{tabular}{ |p{1.2cm}||p{1.5cm}| |p{1.5cm}||p{1.5cm}| |p{2.3cm}||p{2.5cm}| | }
\hline
 Distance&\multicolumn{3}{|c|}{Microwave temperature asymmetry $\frac{\Delta T}{T_{d}}$ ($\mu K$)}&\multirow{2}{*}{Spectral index ($\beta$)}&\multirow{2}{*}{Temperature $T_{d}$ (K)}\\
 \cline{2-4}
  r (kpc) & 857 GHz &545 GHz &353 GHz& & \\
\hline
   15&0.0296&0.351&8.904 &1.822&17.083\\
   20&0.0123&0.205&6.140&1.847&16.845\\
   25&0.0037&0.0586&3.783&1.853&16.848\\
  \hline
  \end{tabular}
}
\end{table}

\par

The optical depths for three frequencies and two models are presented in Tables 12 and 13.
\begin{table}[h!]   
\tbl{M81 optical depth for DL07 model.} 
{
\centering
\begin{tabular}{ |p{2cm}||p{2cm}| |p{2cm}||p{2cm}| }
 \hline
  \multicolumn{4}{|c|}{DL07, $\beta=2$, $T=16.92K$} \\
 \hline
 Distance\newline r (kpc)
 
 &Optical depth\newline$\tau_{857}$ $(10^{-3})$ 
 
 &Optical depth\newline$\tau_{545}$ $(10^{-3})$
 
 &Optical depth\newline$\tau_{353}$ $(10^{-3})$ \\
 
  \hline
  15&$1.57$&$3.48$&$5.80$\\
 
  20&$1.32$&$2.94$&$4.90$\\
 
  25&$1.14$&$2.60$&$4.34$\\
  \hline
  \hline
\multicolumn{4}{|c|}{DL07,  $\beta=2$, $T\neq const$}\\ 
   \hline
Distance\newline r (kpc)

&Optical depth\newline $\tau_{857}$ $(10^{-3})$     
 
&Optical depth\newline $\tau_{545}$ $(10^{-3})$

&Optical depth\newline $\tau_{353}$ $(10^{-3})$\\

  \hline

   15&$1.55$&$3.46$&$5.77$\\
 
   20&$1.32$&$2.94$&$4.94$\\
  
	 25&$1.15$&$2.61$&$4.38$\\
 
	\hline
  \end{tabular} \label{tab:2}}
	
\end{table}

\begin{table}[h!]   
\tbl{M81 optical depth for GNILC  model.}
{
\centering
\begin{tabular}{ |p{2cm}||p{2cm}| |p{2cm}||p{2cm}| }
 \hline
  \multicolumn{4}{|c|}{GNILC, $\beta\neq const2$, $T=const$} \\
  \hline
 Distance\newline r (kpc)
 &Optical depth\newline$\tau_{857}$ $(10^{-2})$
 &Optical depth\newline$\tau_{545}$ $(10^{-2})$
 &Optical depth\newline$\tau_{353}$ $(10^{-2})$\\
  \hline
    
   15&0.531&1.084&1.726\\

   20&0.402&0.848&1.354\\
 
   25&0.360&0.764&1.230\\
  \hline
  \end{tabular} \label{table:3}}
 \end{table}

In the case of Table 12 the optical depth varies significantly in GNILC model, where $\beta\neq const$. In this case again there is almost no temperature dependency, hence the optical depths for DL07 and GNILC models (when both temperature and spectral index are constant) are determined only by the flux density. Thus, the values of optical depths are close to each other. 

\textit{M81 dust rotational velocities and masses}: As in case of M82, we use Eq.(\ref{V}) for calculation of velocities, where for galaxy disk the inclination angle we adopt as $58^{{\circ}}$ (we assume, that the halo has the same inclination angle as the disk). The velocities have been illustrated in Tables 14 and 15. Once again,  we  obtain the most high value at $353 \mathrm{GHz}$ (due to high value of microwave temperature asymmetry). However, these values are significantly low compared to M31’s velocities \cite{Amekhyan} because of relatively low value of optical depth. For example, optical depths for M31 galaxy (at $353 \mathrm{GHz}$) vary within 10$^{-3}-10^{-4}$ range, while in this case the order of magnitude of the optical depth is more or less the same, similarly to the discussion in the previous section. 
\begin{table}[h!]   
\tbl{M81 dust rotational velocities for DL07 model obtained from microwave temperature asymmetry data.}
{
\centering
\begin{tabular}{ |p{2cm}||p{2cm}||p{2cm}||p{2cm}|}
\hline
\multicolumn{4}{|c|}{DL07, $\beta=const$,    $T=const$} \\
\hline
Distance\newline r (kpc)
&$\nu=857$ GHz\newline  $V_{rot}$ (km/s)
&$\nu=545$ GHz\newline  $V_{rot}$ (km/s)
&$\nu=353$ GHz\newline  $V_{rot}$ (km/s)\\
\hline
  15& 3.35&81.2&272.37 \\

  20&1.66&12.3&222.16\\
 
  25&0.57&4&154.53 \\

\hline
\hline

  \end{tabular} \label{table:7}}

\end{table}

\begin{table}[h!]   
\tbl{M81 dust rotational velocities for GNILC model obtained from microwave temperature asymmetry data.}
{
\centering
\begin{tabular}{ |p{2cm}||p{2cm}||p{2cm}||p{2cm}|}
 \hline
  \multicolumn{4}{|c|}{GNILC, $\beta\neq const$,   $T=const$} \\
 \hline
Distance\newline r (kpc)
&$\nu=857$ GHz\newline  $V_{rot}$ (km/s)
&$\nu=545$ GHz\newline  $V_{rot}$ (km/s)
&$\nu=353$ GHz\newline  $V_{rot}$ (km/s)\\
  \hline
   15&0.98& 13.21& 91.92\\

   20&0.54& 9.90 &80.75\\
 
   25& 0.18 & 3.15&54.93\\

  \hline
\hline
 
  \end{tabular} \label{table:8}}
\end{table}
The dust rotation velocity can be obtained from the dust mass as well. There are various estimates of M81’s dust mass. According to \cite{ Bendo} the mass is $3.4\times 10^{7}\mathrm{M_{\odot}}$ for $100-500\mathrm{\mu m}$ wavelengths range. Another value of dust mass, $1.293\times 10^{7} \mathrm{M_{\odot}}$, is given in DustPedia \cite{DustPedia}.  
Here we also use Eq.(\ref{H}) for determining dust masses at three frequencies and two dust models. Since the mass absorption coefficient ($k$) depends only on given frequency, we adopt the same $k$ which we have used for M82. As for the optical depth, we use values from Tables 12 and 13, calculated from MMB spectrum. 

Performing the same analysis as of M82, we find out the masses and the velocities for M81, as presented in Tables 16-19. 

\begin{table}[h!]   
\tbl{M81 dust mass for DL07 model.}
{
\centering
\begin{tabular}{ |p{2cm}||p{2cm}||p{2cm}||p{2cm}|}
 \hline
  \multicolumn{4}{|c|}{DL07, $\beta=const$,   $T=const$} \\
 \hline
Distance\newline r (kpc)
&$\nu=857$ GHz\newline $M_{\odot}$ $(10^{6})$
&$\nu=545$ GHz\newline $(M_{\odot}$ $(10^{6})$
&$\nu=353$ GHz\newline $M_{\odot}$ $(10^{6})$\\ 
\hline
   15& 0.260& 1.210 & 4.104 \\

   20&0.219 & 1.025 &3.469 \\
 
   25& 0.189 & 0.904&3.073 \\
 \hline
\hline
\end{tabular} \label{table:16}}
\end{table}

\begin{table}[h!]   
\tbl{M81 dust mass for GNILC model.}
{
\centering
\begin{tabular}{ |p{2cm}||p{2cm}||p{2cm}||p{2cm}|}
 \hline
  \multicolumn{4}{|c|}{GNILC, $\beta\neq const$,   $T=const$} \\
 \hline
Distance\newline r (kpc)
&$\nu=857$ GHz\newline  $M_{\odot}$ $(10^{5})$
&$\nu=545$ GHz\newline  $M_{\odot}$ $(10^{6})$
&$\nu=353$ GHz\newline  $M_{\odot}$ $(10^{6})$\\ 
  \hline
   15& 8.814& 3.770 & 9.147 \\

   20&6.673 & 2.949 &8.575 \\
 
   25& 5.976 & 2.657  &8.199 \\

  \hline
\hline
 
  \end{tabular} \label{table:17}}
\end{table}

\begin{table}[h!]   
\tbl{M81 dust rotational velocities for DL07 model derived with Eq.(3).}
{
\centering
\begin{tabular}{ |p{2cm}||p{2cm}||p{2cm}||p{2cm}|}
 \hline
  \multicolumn{4}{|c|}{DL07, $\beta=const$,   $T=const$} \\
 \hline
Distance\newline r (kpc)
&$\nu=857$ GHz\newline  $V_{rot}$ (km/s)
&$\nu=545$ GHz\newline  $V_{rot}$ (km/s)
&$\nu=353$ GHz\newline  $V_{rot}$ (km/s)\\ 
  \hline
   15& 0.273& 0.591& 1.088 \\

   20&0.251 & 0.544 &1.00\\
 
   25& 0.233 & 0.510  &0.941 \\

  \hline
\hline
 
  \end{tabular} \label{table:18}}
\end{table}

\begin{table}[h!]   
\tbl{M81 dust rotational velocities for GNILC model derived with Eq.(3).}
{
\centering
\begin{tabular}{ |p{2cm}||p{2cm}||p{2cm}||p{2cm}|}
\hline
\multicolumn{4}{|c|}{GNILC, $\beta\neq const$,   $T=const$} \\
\hline
 
Distance\newline r (kpc)
&$\nu=857$ GHz \newline  $V_{rot}$ (km/s)
&$\nu=545$ GHz\newline  $V_{rot}$ (km/s)
&$\nu=353$ GHz\newline  $V_{rot}$ (km/s)\\ 
  \hline
   15& 0.504& 1.043 & 1.625 \\

   20&0.438 & 0.922 &1.573 \\
 
   25& 0.415 & 0.875  &1.538 \\
 
\hline
\hline
\end{tabular} \label{table:19}}
\end{table}

Thus, we use two approaches for deriving the dust rotation velocity. First, we compute it using the microwave temperature asymmetry data. Then, we obtain the velocity via Eq.(3) and find that the obtained velocities in the first case are significantly higher than those determined by microwave temperature  asymmetry data. In the second case i.e. where we use pure dust parameters without microwave data, the velocities have smaller values.

\section{Gravity and dark sector}

The essential fraction of the dark matter in galaxies is commonly believed to be stored in the halos and various density profiles are proposed to fit the observations. Here we list some of the commonly used profiles to model dark matter halos. The so-called pseudo-isothermal profile is written as \cite{PIS}
\begin{equation}\label{PIS}
\rho (r)=\rho _{0}\left[1+\left({\frac {r}{r_{c}}}\right)^{2}\right]^{-1},
\end{equation} 
where $\rho _{0}$ denotes the central density and $r_{c}$ is the core radius. The Navarro-Frenk-White (NFW) \cite{NFW} profile with broad range of applications is
\begin{equation}\label{NFW}
\rho (r)={\frac {\rho _{crit}\delta _{c}}{\left({\frac {r}{r_{c}}}\right)\left(1+{\frac {r}{r_{c}}}\right)^{2}}},
\end{equation} 
where $\rho _{crit}$ is the critical density of the universe defined as
\begin{equation}
\rho _{crit}= \frac{3 H^2}{8 \pi G},
\end{equation}
$H$ stands for the Hubble constant. One of important features of this profile is that it depends directly on the cosmological parameter $\rho _{crit}$. The dimensionless parameter $\delta _{c}$ relates the two radii i.e. $\rho _{crit}$ to $\rho _{0}$ for the halo under consideration: $\rho _{crit}= \delta_{c} \rho_{0}$. Along with NFW, other profiles are also considered for modeling the halos, e.g. the Burkert one\cite{Burkert} 
\begin{equation}\label{Burkert}
\rho (r)={\frac {\rho _{crit}\delta _{c}}{\left({1+(\frac {r}{r_{c}}})\right)\left(1+{(\frac {r}{r_{c}})}^2\right)}}
\end{equation} and Moore profile \cite{Moore}:
\begin{equation}\label{Moore}
\rho (r)={\frac {\rho _{crit}\delta _{c}}{\left({\frac {r}{r_{c}}}\right)^{\frac{3}{2}}\left(1+{(\frac {r}{r_{c}})}^{\frac{3}{2}}\right)}}
\end{equation}
These three profiles have been used to model the dark matter halo of M31 \cite{M31}. 

Recalling the fact that the first indirect observation of dark matter in galaxies was related to the virial theorem
\begin{equation}\label{Vir}
\sigma^2 = \frac{GM}{R},
\end{equation}
where $\sigma$, $M$ and $R$ stand for the velocity dispersion, the virial mass and the virial radius, respectively, for above mentioned  halo profiles the velocity dispersions read as (see \cite{Amekhyan} for more details)
\begin{equation}\label{VNFW}
\sigma^2_{NFW}= 4\pi G\rho_c \frac{r_c^3}{r}(ln\big(1+\frac{r}{r_c}\big)-
\frac{\frac{r}{r_c}}{(1+\frac{r}{r_c})}),
\end{equation}
\begin{equation}\label{VMoore}
\sigma^2_{Moore}= \frac{8}{3}\pi G \rho_c\frac{r_c^3}{r}(ln(1+\big(\frac{r}{r_c}\big)^{\frac{3}{2}})), 
\end{equation}
\begin{equation}\label{VBurkert}
\sigma^2_{Burkert}= 2\pi G\rho_c \frac{r_c^3}{r} ((ln\big(1+\frac{r}{r_c}\big) \sqrt{1+\big(\frac{r}{r_c}\big)^2}) -\arctan(\frac{r}{r_c})).
\end{equation}

In addition to these profiles, another widely used profile is the Einasto law  \cite{Ein}
\begin{equation}\label{Ein}
\rho (r)=\rho _{e}e^{\left(-d_{n}\left(\left({\frac {r}{r_{e}}}\right)^{\frac {1}{n}}-1\right)\right)}
\end{equation}
In Einasto model \cite{EinM}  $d_{n}$ is function of $n$ such that $\rho _{e}$ is the density at the radius  $r_{e}$ of the volume containing half of the total mass.

The cosmological constant $\Lambda$ entering the Einstein equations 
\begin{equation}\label{L}
G_{\mu \nu} + \Lambda g_{\mu \nu}= \frac{8 \pi G}{c^4} T_{\mu \nu},
\end{equation}
is currently considered to fit the dark energy data.

Various classes of modified gravity models are being considered to explain the dark matter and dark energy data. The models include e.g. scalar field entering the action 
\begin{equation}\label{BD}
S={\frac {c^4}{16\pi }}\int \;\left(\phi R-{\frac {\omega }{\phi }}\partial _{a}\phi \partial ^{a}\phi \right){\sqrt {-g}}d^{4}x,
\end{equation}
where $\omega$ is a dimensionless constant known as Dicke coupling constant \cite{BC}. Other models are based on the modification of Einstein-Hilbert action
\begin{equation}\label{EH}
S={c^4 \over 16 \pi G }\int R{\sqrt {-g}}\, d ^{4}x,
\end{equation}
where $R$ is the Ricci scalar. This defines the $f(R)$ theories of gravity \cite{Far, Cap}. All modified theories are reduced to ordinary General Relativity at specific regimes. 

The Modified Newtonian Dynamics (MOND) \cite{M1} is one of the well known model for explaining the dark matter problem. In MOND the Newton's second law is modified as
\begin{equation}\label{M}
F= m a \mu (\frac{a}{a_0}),
\end{equation}
where $\mu (\frac{a}{a_0})$ is the ``extrapolating function" and depends on a parameter $a_0 \approx 1.2 \times 10^{-12} ms^{-2}$. Thus, although for $a_0 << a$ Newton's second law remains valid, in the so-called ``deep-MOND" regime, it is modified to
\begin{equation}\label{dM}
F= m \frac{a^2}{a_0}, \quad (a<< a_0).
\end{equation}  
Consequently, for the circular motion of an object with mass $m$ around another object with mass $M$ one has
\begin{equation}\label{MG}
\frac{GmM}{r^2} = m \frac{(\frac{v^2}{r})^2}{a_0}.
\end{equation}
Considering the above relation, it turns out that one can explain the so-called ``flat rotation curves" of galaxies without any need of dark matter. Meantime, it is possible to interpret Eq.(\ref{MG}) as the modification of Newtonian gravity, leaving Newton's law intact. In such case, the modified gravitational potential is
\begin{equation}\label{MP}
\Phi= (GMa_0)^{\frac{1}{2}} ln r.
\end{equation}
Thus, MOND proposes a modification of gravity according to Eq.(\ref{MG}) without any further need of dark matter. For M31, particularly, the rotation curves predicted by MOND seem to be inconsistent with observations \cite{M31-3}. 
 
We mention one more approach to describe the dark matter and dark energy within a single concept. Namely, considering the Newton's theorem on the equivalency of gravitational fields produced by sphere and that of the point, the weak-field limit of General Relativity is written as \cite{VG,GS1}
\begin{equation}\label{WF}
g_{00} = 1 - \frac{2 G m}{r c^2} - \frac{\Lambda r^2}{3};\,\,\, g_{rr} = (1 - \frac{2 G m}{r c^2} - \frac {\Lambda r^2}{3})^{-1}.
\end{equation}
Then the cosmological constant enters naturally in the gravity equations (both General Relativity and Newtonian one as its weak-field limit) which in its turn enables one to study the effect of $\Lambda$ not only in the cosmological scales, but also in the Local Supercluster scale \cite{VG,GS4}. Consequently, the virial theorem is written as
\begin{equation}\label{virialL}
\sigma^2 = \frac{GM}{R} + \frac{\Lambda c^2 R^2}{6},
\end{equation}
while for rotational velocity we have
\begin{equation}\label{NVL}
V(r)^2=\frac{GM_{d}}{r}-\frac{\Lambda c^2 r^2 }{3}.
\end{equation}
This relation can be considered as further support to the conclusions in the previous sections, but now from entirely different point of view. Namely, as shown in \cite{VG,GS1,GS4} Eqs(22),(23) are able to describe the dynamics of galactic halos and at higher scale galaxy configurations. However, if considering the dust mass obtained above as significant component of halos, one comes at contradiction i.e. the square of rotational velocity, $V^2$, according to Eq.(23) will decrease and even can become negative. That contradiction can either indicate the limitations of the model on that scales or that the actual contribution of dust in the halos has to be small.

\section{Conclusion}

We studied the thermal dust contribution in the M81 and M82 galactic halos' rotation, up to $25\mathrm{kpc}$. We found dust rotation velocities with three frequency bands and two dust models, using Doppler induced microwave temperature asymmetry data. We estimated the dust rotation velocities via dust mass, which again depends on the given frequency.  We obtained the lower values of the dust opacity using Doppler formula for microwave temperature asymmetry and conclude that velocities obtained via microwave asymmetry data (with Doppler effect) has a higher value at $353 \mathrm{GHz}$ band, with $V=160-80\mathrm{km/s}$ for DL07 dust model and $V=14-10\mathrm{km/s}$ for GNILC model (both for M82 galaxy). Since the dust spectral index specifies the dust model, we considered DL07 model with $\beta=2$ spectral index and GNILC of varying spectral index; the dust characteristic parameters significantly different for those models, hence vary the values of the dust velocities. In fact, spectral index is a crucial parameter through which almost all properties of the dust are being fixed. Then we derived the rotation velocities of dust via Eq.(2) and obtained a significantly lower values $3.5-1.5\mathrm{km/s}$ at $353 \mathrm{GHz}$.

We also analyzed the relevance of the obtained results to certain modified gravity models proposed to explain the dark matter indications including the flat rotation curves in galaxies. We mentioned a self-consistent approach to the dark matter problem based on Newton's theorem on sphere-point equivalency, where the cosmological constant $\Lambda$ is introduced naturally in the weak-field General Relativity. The notable conclusion is that although we used different models, we arrived at the same conclusion on the minor role of the dust in the entire dynamical mass of the considered galaxies.

\section{Acknowledgments}

We are thankful to the referees for valuable comments and to A. Kashin, S. Mirzoyan and H. Khachatryan for discussions and help. The use of  Planck data in the Legacy Archive for Microwave Background Data Analysis (LAMBDA) and HEALPix \cite{Gor} package is acknowledged. 
AS acknowledges the partial support by the ICTP through AF-04.


\end{document}